# A Magnetic Neuroevolution Potential for Large Scale Spin–Lattice Simulations

Rongkun Chen[1], Junhan Li[1], Feng Xiao[2], Shiqian Hu[1*]

1. School of Physics and Astronomy, Yunnan Key Laboratory for Quantum Information, Yunnan University, Kunming 650091, People's Republic of China

2. Spallation Neutron Source Science Center, Dongguan 523803, China

[*]Corresponding author:

shiqian@ynu.edu.cn

Accurately describing magnetic order, lattice vibrations, and their coupling at finite temperatures within a unified framework remains a major challenge in atomistic simulations of magnetic materials. In this work, we develop a magnetic neuroevolution potential that explicitly incorporates magnetic degrees of freedom, establishing a potential-energy description jointly determined by atomic structures and magnetic configurations and enabling unified predictions of atomic forces and magnetic energy gradients. Using CrN, Fe, and FeGe with collinear or noncollinear magnetic configurations as representative systems, we demonstrate that the model outperforms standard neuroevolution potentials without magnetic descriptors in predicting energies, atomic forces, and virials, while yielding equilibrium lattice constants consistent with experimental values. Phonon calculations for different magnetic states show that the model captures lattice vibrational responses to changes in magnetic order. Finite-temperature spin-lattice dynamics simulations further reveal the temperature dependence of magnetic excitation spectra, while calculations of the average atomic magnetic moment as a function of temperature demonstrate the influence of spin-lattice coupling on magnetic transition temperatures and convergence with respect to system size. For Fe, explicitly including spin-lattice coupling brings the predicted magnetic transition temperature closer to experiment, whereas the corresponding differences in FeGe are smaller. Performance benchmarks demonstrate that the method enables coupled dynamics simulations of systems containing more than two million atoms on a single consumer-grade GPU. By combining magnetic-configuration resolution with large-scale computational efficiency, the framework provides an efficient tool for investigating finite-temperature dynamics and spin-lattice coupling in large-scale magnetic systems.

## INTRODUCTION

Magnetic materials have long attracted widespread interest owing to their rich physical and chemical properties. For example, B20-type FeGe hosts skyrmion spin textures and exhibits distinctive chiral phenomena associated with magnetism and phonons[1–4]. These materials exhibit a rich variety of magnetically ordered states and magnetic-order-dependent structural and dynamical behavior, providing an important platform for condensed matter research and functional device design. Accurately describing these phenomena requires simultaneous consideration of atomic structures, magnetic degrees of freedom, and their interactions[5–7]. In addition, magnetic materials have important applications in information storage, sensing devices, and biomedicine[8–10]. Therefore, developing theoretical methods that combine accuracy with computational efficiency is essential for understanding the finite-temperature properties and spin-lattice dynamics of magnetic materials, and is of considerable significance for both fundamental research and practical applications[11,12].

First-principles calculations based on density functional theory (DFT) are an important tool for investigating magnetic materials. Combined with approaches such as the four-state method and energy mapping[13–15], DFT calculations enable the extraction of magnetic interaction parameters and further investigation of magnetic properties. However, DFT calculations for magnetic systems are typically computationally demanding, limiting both the accessible system sizes and the number of magnetic configurations that can be sampled. Consequently, relying solely on DFT remains challenging for studies requiring extensive configurational sampling, such as the finite-temperature evolution of magnon spectra and coupled spin-lattice dynamics.

In recent years, machine-learning interatomic potentials have developed rapidly, enabling efficient atomistic simulations with accuracy approaching that of their first-principles reference data and providing new computational approaches for condensed matter physics and materials science[16]. These methods are typically trained on atomic configurations and their corresponding energies and atomic forces to learn the potential energy surface (PES), thereby enabling simulations of structural and dynamical behavior under different conditions, including temperature and strain. However, conventional machine-learning potentials generally use only atomic species and coordinates as inputs, without explicitly incorporating magnetic degrees of freedom[17,18]. They therefore lack the information needed to distinguish different magnetic configurations for a given atomic structure, making it difficult to provide a unified description of the potential energy surfaces associated with different magnetic states or to directly predict magnetic-order-dependent phonon responses, magnetic transition temperatures, and spin dynamics. To overcome these limitations, several machine-learning approaches for magnetic systems have been proposed, including magnetic moment tensor potentials (mMTPs)[18], magnetic high-dimensional neural network potentials (mHDNNPs)[17], magnetoelastic machine-learning interatomic potentials (ML-IAPs)[19], DeepSPIN[20], and SpinGNN[21]. These approaches have extended the capabilities of machine-learning potentials for describing magnetic systems to varying degrees and have advanced large-scale spin-lattice simulations. For example, SpinGNN has enabled million-atom simulations using multiple GPUs in parallel[21]. Nevertheless, performing large-scale spin-lattice simulations over long timescales with limited computational resources places considerable demands on model efficiency. Achieving both accurate magnetic descriptions and computational efficiency in large-scale simulations therefore remains an important research challenge.

In this work, we propose a magnetic machine-learning potential framework based on the neuroevolution potential (NEP)[22,23]. By combining physically interpretable magnetic descriptors with a compact feedforward neural network, the framework enables an accurate description of the potential energy surfaces of magnetic materials. When trained and evaluated on the same datasets, the proposed framework reduces test-set energy prediction errors by up to an order of magnitude relative to the original NEP without magnetic descriptors. Using CrN, Fe, and FeGe as representative systems, we validate its ability to capture magnetic-order-dependent phonon responses through phonon calculations for different magnetic states. We further assess its predictive capability for magnetic properties by calculating magnon spectra and magnetic transition temperatures. Finally, we benchmark its computational performance for Fe and FeGe systems of varying sizes on a single consumer-grade RTX 4090 GPU. The results demonstrate that the framework supports efficient simulations of magnetic systems containing two million atoms on a single GPU, providing an accurate and efficient computational tool for large-scale coupled spin-lattice simulations and investigations of finite-temperature properties in magnetic materials.

## Methods

### Standard NEP Framework

Following the standard NEP framework[22,23], the total potential energy is represented as a sum of local atomic contributions:

$$U = \sum_i U_i, U_i = N_{t_i}(\hat{q}_i). \quad (1)$$

Here, $t_i$ denotes the chemical species of atom $i$, $N_{t_i}$ is the corresponding species-dependent neural network, and $\hat{q}_i$ is a descriptor vector characterizing its local atomic environment. The descriptors are constructed from relative atomic positions and sums over neighbors, with rotational invariance enforced through scalar contractions. This construction ensures invariance under translations, rotations, and permutations of identical atoms.

The local distribution of interatomic distances is encoded using trainable radial functions:

$$g_{n,t_it_j}^{(a)}(\mathbf{r}) = \sum_{k=0}^{N^a} c_{nk,t_it_j}^{(a)} f_k^{(a)}(\mathbf{r}), a \in \{\hat{R}, \hat{A}\}. \quad (2)$$

The labels $\hat{R}$ and $\hat{A}$ distinguish the radial functions used in the radial and angular descriptor channels, respectively. The expansion coefficients $c_{nk,t_it_j}^{(a)}$ are trainable and depend on the species pair. The fixed basis functions combine Chebyshev polynomials of the first kind with a cutoff function:

$$f_k^{(a)}(\mathbf{r}) = \frac{1}{2}[T_k(x_a(\mathbf{r})) + 1]f_c^{(a)}(\mathbf{r}), \quad (3)$$

$$x_a(\mathbf{r}) = 2(\frac{\mathbf{r}}{r_c^{(a)}} - 1)^2 - 1, \quad (4)$$

$$f_c^{(a)}(\mathbf{r}) = \begin{cases} \frac{1}{2}[\cos(\frac{\pi\mathbf{r}}{r_c^{(a)}} + 1)], 0 \le \mathbf{r} < r_c^{(a)} \\ 0, otherwise \end{cases} \quad (5)$$

Here, $r_c^{(a)}$ is the corresponding cutoff radius. Both the cutoff function and its first derivative vanish at the cutoff, ensuring that the basis functions and their first derivatives approach zero continuously. The radial two-body descriptors are obtained by summing over neighboring atoms:

$$\hat{q}_{n,i}^{(\hat{R})} = \sum_{j \neq i}^{N^{(\hat{R})}} g_{n,t_i t_j}^{(\hat{R})}(\mathbf{r}_{ij}). \quad (6)$$

Here, $\mathbf{r}_{ij} = \mathbf{R}_j - \mathbf{R}_i$, $r_{ij} = |\mathbf{r}_{ij}|$, and sums over all the radial neighbor set. These channels characterize the neighbor-distance distribution through distinct radial weighting functions, while the species-pair-dependent coefficients distinguish different chemical environments.

Interatomic distances alone cannot fully distinguish different coordination geometries. NEP therefore introduces angular descriptors by projecting the neighbor-direction distribution onto spherical harmonics:

$$s_{nlm,i} = \sum_{j \neq i}^{N^{(\hat{A})}} g_n(\mathbf{r}_{ij}) Y_{lm}(\mathbf{e}_{ij}), \mathbf{e}_{ij} = \frac{\mathbf{r}_{ij}}{r_{ij}}. \quad (7)$$

Rotationally invariant three-body descriptors are obtained by contracting the spherical-harmonic components:

$$\hat{q}_{nl,i}^{(\hat{A})} = \sum_{m=-l}^{l} \left| s_{nlm,i} \right|^2. \quad (8)$$

Using the spherical-harmonic addition theorem, this expression can be rewritten as

$$\hat{q}_{nl,i}^{(\hat{A})} = \frac{2l+1}{4\pi} \sum_{j,k \neq i}^{N^{(\hat{A})}} g_n(\mathbf{r}_{ij}) g_n(\mathbf{r}_{ik}) P_l(\mathbf{e}_{ij} \cdot \mathbf{e}_{ik}). \quad (9)$$

Here, both neighbor indices run over $N^{(\hat{A})}$, and $P_l(\mathbf{e}_{ij} \cdot \mathbf{e}_{ik})$ is a Legendre polynomial. The angular descriptors encode both interatomic distances and bond angles. Their spherical-harmonic representation allows the intermediate quantities to be accumulated through a single sum over neighbors, avoiding an explicit double sum over neighbor pairs. Additional higher-body structural invariants, if enabled, can be appended without altering the magnetic extension described below.

The radial and angular descriptors are concatenated into a structural descriptor vector:

$$\hat{q}_i^{\mathrm{str}} = \hat{q}_{n,i}^{(\hat{R})} \oplus \hat{q}_{nl,i}^{(\hat{A})}, \tag{10}$$

which is passed to a species-dependent feedforward neural network with a single hidden layer:

$$U_i = \sum_{\mu=1}^{N_{\mathrm{neu}}} w_{t_i,\mu}^{(1)} \tanh\left[\sum_{d=1}^{\mathrm{dim}} w_{t_i,\mu d}^{(0)}\, \hat{q}_{d,i} - b_{t_i,\mu}^{(0)}\right] - b_{t_i}^{(1)}. \tag{11}$$

Here, $\hat{q}_{d,i}$ denotes a scaled descriptor component and $N_{\mathrm{neu}}$ is the number of hidden neurons. Atomic forces are obtained analytically from the same total energy function:

$$\mathbf{F}_i = -\nabla_{\mathbf{r}_i} U = -\sum_j \sum_{d=1}^{\mathrm{dim}} \frac{\partial U_j}{\partial \hat{q}_{d,j}} \frac{\partial \hat{q}_{d,j}}{\partial \mathbf{R}_i}. \tag{12}$$

This construction separates local environment representation from energy regression: the descriptors encode structural information and symmetry constraints, while the neural network provides the nonlinear mapping to atomic energies. Because standard NEP descriptors depend only on atomic positions and chemical species, they cannot distinguish different magnetic configurations for a given atomic structure. In the following section, explicit spin-dependent channels are introduced by reusing the radial functions of the angular descriptors.

**Magnetic Descriptor Design**

To incorporate magnetic degrees of freedom, each magnetically active atom is assigned a Cartesian magnetic-moment vector $\mathbf{S}_i$, with magnitude $s_i = |\mathbf{S}_i|$. The total energy is then expressed as a function of both atomic positions and magnetic moments:

$$U(\{\mathbf{R}_i\},\{\mathbf{S}_i\}) = \sum_i^{N_{t_i}} (\hat{q}_i^{\mathrm{str}} \oplus \hat{q}_i^{\mathrm{spin}}). \tag{13}$$

Treating atomic positions and magnetic moments as joint variables of the potential-energy surface follows the general strategy of magnetic extensions of atomistic energy representations[18,24].

The magnetic descriptor vector comprises five blocks: isotropic exchange (Sex), bond-dependent anisotropic exchange (San), a chiral exchange channel (Sdm), single-ion anisotropy (Sia), and on-site moment magnitude (Smag). The neighbor-dependent blocks share the angular-channel radial functions. For compactness, we define

$$h_{n,ij} = g_{n,t_i t_j}^{(\hat{A})}(\mathbf{r}_{ij}), \quad a_{ij} = \mathbf{S}_i \cdot \mathbf{e}_{ij}, \quad b_{ij} = \mathbf{S}_j \cdot \mathbf{e}_{ij}. \tag{14}$$

The weight $h_{n,ij}$ includes the cutoff function defined in Section 2.1; no additional cutoff factor is applied. Unless otherwise specified, the neighbor sums below run over $N^{(\hat{A})}$.

The isotropic exchange block is constructed from pairwise spin scalar products:

$$\text{Sex}: \hat{q}_{n,i}^{\mathrm{Sex}} = \sum_{j\neq i} h_{n,ij} \, (\mathbf{S}_i \cdot \mathbf{S}_j). \tag{15}$$

This block encodes Heisenberg-like spin correlations, while its radial channels resolve their distance dependence. The anisotropic exchange block additionally couples the spin orientations to the bond direction:

$$\text{San}: \hat{q}_{n,i}^{\mathrm{San}} = \sum_{j\neq i} h_{n,ij} \, (\mathbf{S}_i \cdot \mathbf{e}_{ij})(\mathbf{S}_j \cdot \mathbf{e}_{ij}) = \sum_{j\neq i} h_{n,ij} \, a_{ij} b_{ij}. \tag{16}$$

Together, “Sex” and “San” provide the angular structures associated with isotropic exchange and dipolar-like interactions. Their finite-cutoff form does not, by itself, constitute a full treatment of long-range magnetostatic interactions.

The chiral exchange block is defined using a scalar triple product:

$$\mathrm{Sdm}\colon \hat{q}_{n,i}^{\mathrm{Sdm}} = \sum\nolimits_{j\neq i} h_{n,ij}\, \mathbf{e}_{ij} \cdot (\mathbf{S}_i \times \mathbf{S}_j). \tag{17}$$

This channel resolves the bond-parallel component of Dzyaloshinskii-Moriya interactions[25,26]. It is invariant under simultaneous proper rotations of positions and spins and is even under global spin reversal, but changes sign under spatial inversion; this sign change is the origin of chirality. Its form should therefore be distinguished from a general environment-dependent Dzyaloshinskii-Moriya vector.

Motivated by bond-direction-dependent local anisotropy models[27], the single-ion anisotropy block couples the central moment quadratically to the local bond geometry:

$$\begin{aligned}\mathrm{Sia}\colon \hat{q}_{n,i}^{\mathrm{Sia}} &= \sqrt{\frac{3}{4\pi}} \sum\nolimits_{j\neq i} h_{n,ij} \left[\frac{3}{2}(\mathbf{S}_i \cdot \mathbf{e}_{ij})^2 - \frac{1}{2}|\mathbf{S}_i|^2\right] \\ &= \sqrt{\frac{3}{4\pi}} \sum\nolimits_{j\neq i} h_{n,ij} \left[\frac{3}{2} a_{ij}^2 - \frac{1}{2} s_i^2\right].\end{aligned} \tag{18}$$

Its physical meaning can be made explicit by defining a local traceless environment tensor:

$$\widehat{\mathbf{Q}}_{n,i} = \sum\nolimits_{j\neq i} h_{n,ij} \left[\frac{3}{2} \mathbf{e}_{ij} \otimes \mathbf{e}_{ij} - \frac{1}{2}\mathbf{I}\right], \tag{19}$$

$$\hat{q}_{n,i}^{\mathrm{Sia}} = \mathbf{S}_i \widehat{\mathbf{Q}}_{n,i} \mathbf{S}_i. \tag{20}$$

Here, $\mathbf{I}$ is the identity tensor. For a nonzero local moment, each bond contribution is proportional to $s_i^2 P_2(\cos\theta_{ij})$, where $\theta_{ij}$ is the angle between $\mathbf{S}_i$ and $\mathbf{e}_{ij}$. This block characterizes the quadrupolar alignment of the local moment with its environment. Unlike the former linear spin-environment channel, the quadratic form is even under spin reversal by construction. It provides a rank-2 anisotropy channel rather than an explicit general rank-4 representation of cubic anisotropy.

The on-site block describes energy variations associated with the local moment magnitude:

$$\text{Smag}: \hat{q}_{n,i}^{\text{Smag}} = u_i^{n+1}, u_i = \eta s_i, \tag{21}$$

where $\eta$ is the multiplicative scaling factor stored as "spin_mag_scale". If unspecified, $\eta$ is set to the inverse of the maximum moment magnitude in the training data. These channels depend only on the central moment and have no explicit positional dependence. They therefore contribute directly to the magnetic energy gradient but not to positional descriptor derivatives. Because the first channel is linear in $|\mathbf{S}_i|$, its derivative is undefined at $\mathbf{S}_i = 0$, and this limit requires separate treatment. A magnitude-dependent energy representation should also be distinguished from longitudinal spin dynamics[28]: a length-preserving spin integrator does not evolve $s_i$.

The magnetic blocks are concatenated with the structural descriptors to form the neural network input:

$$\hat{q}_i = \hat{q}_i^{\text{str}} \oplus \hat{q}_{n,i}^{\text{Sex}} \oplus \hat{q}_{n,i}^{\text{San}} \oplus \hat{q}_{n,i}^{\text{Sdm}} \oplus \hat{q}_{n,i}^{\text{Sia}} \oplus \hat{q}_{n,i}^{\text{Smag}}. \tag{22}$$

All five magnetic blocks are invariant under simultaneous proper rotations and even under global spin reversal. Consequently, the learned energy preserves these symmetries in the absence of explicit time-reversal-breaking terms. "Sdm" is a spatial pseudoscalar, whereas the other four blocks are spatial scalars; inclusion of "Sdm" in an unconstrained network does not, on its own, guarantee invariance under improper rotations.

Magnetic energy gradients are derived consistently from the similar scalar energy:

$$\mathbf{w}_i = -\nabla_{\mathbf{S}_i} U = -\sum_{j \neq i} \sum_{d}^{\dim} \frac{\partial U_j}{\partial \hat{q}_{n,j}^{d}} \frac{\partial \hat{q}_{n,i}^{d}}{\partial \mathbf{S}_i}. \tag{23}$$

The derivative includes contributions from descriptors centered on atom i and from neighboring atoms whose descriptors contain $\mathbf{S}_i$. Here, $\mathbf{w}_i$ denotes the generalized magnetic force, which should be distinguished from the spin torque proportional to $\mathbf{S}_i \times \mathbf{w}_i$. Proper-rotation invariance of the energy ensures that the resulting force and magnetic-gradient vectors transform covariantly under simultaneous rotations. The dependence of the descriptors on both positions and spins provides an energy-consistent description of spin-lattice coupling.

Reusing the angular-channel radial functions avoids introducing independent magnetic radial-expansion coefficients. The additional descriptor channels nevertheless increase the number of neural network input-layer weights. Since the neighbor-dependent magnetic blocks involve only single sums over neighbors and Smag is evaluated locally, the extension retains a local computational structure suitable for large-scale simulations.

## Training Procedure and Accuracy Assessment

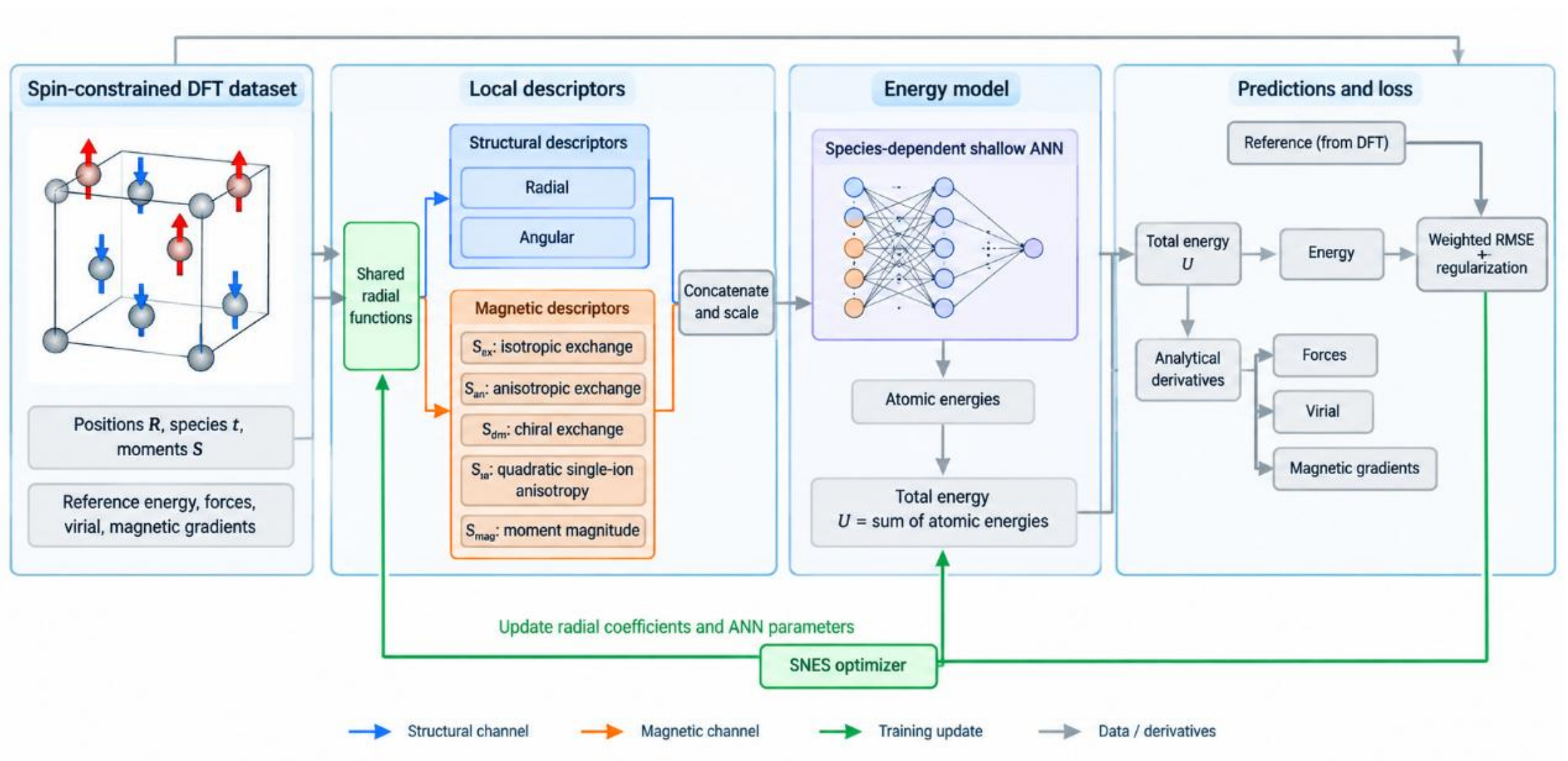

**Fig1. Training workflow of the magnetic neuroevolution potential.** Atomic positions, chemical species, and magnetic moments from spin-constrained DFT calculations are used to construct structural and magnetic descriptors. The magnetic descriptors comprise isotropic exchange (Sex), anisotropic exchange (San), chiral exchange (Sdm), quadratic single-ion anisotropy (Sia), and on-site moment-magnitude (Smag) channels. The neighbor-dependent magnetic channels share the radial functions used by the angular structural descriptors. The concatenated and scaled descriptors are passed to species-dependent feedforward neural networks to obtain atomic energies, which are summed to yield the total energy. Atomic forces, virials, and magnetic energy gradients are evaluated through analytical differentiation. The radial expansion coefficients and neural network parameters are optimized using the separable natural evolution strategy (SNES) by minimizing a weighted loss comprising root-mean-square errors relative to the DFT reference data and regularization terms.

Fig. 1. illustrates the training workflow of the magnetic neuroevolution potential. Atomic positions, chemical species, and magnetic moments from spin-constrained DFT calculations serve as inputs for constructing structural and magnetic descriptors. The structural descriptors encode radial and angular information, while the magnetic descriptors comprise isotropic exchange (Sex), anisotropic exchange (San), chiral exchange (Sdm), quadratic single-ion anisotropy (Sia), and on-site moment-magnitude (Smag) channels. The neighbor-dependent magnetic channels share the radial functions used by the angular structural descriptors. These descriptors are concatenated, scaled, and passed to species-dependent feedforward neural networks to predict atomic energies, which are summed to obtain the total energy.

The training loop shown in Fig. 1 compares the predicted energies, atomic forces, virials, and magnetic energy gradients with their DFT reference values. The derivative quantities are obtained analytically from the same energy model, ensuring consistency between the energy and its derivatives. A weighted loss combines the root-mean-square errors of the predicted quantities with regularization terms. The separable natural evolution strategy (SNES)[29] iteratively updates both the shared radial expansion coefficients and the neural network parameters to minimize this loss, allowing the model to learn the joint dependence of the energy on atomic structures and magnetic configurations.

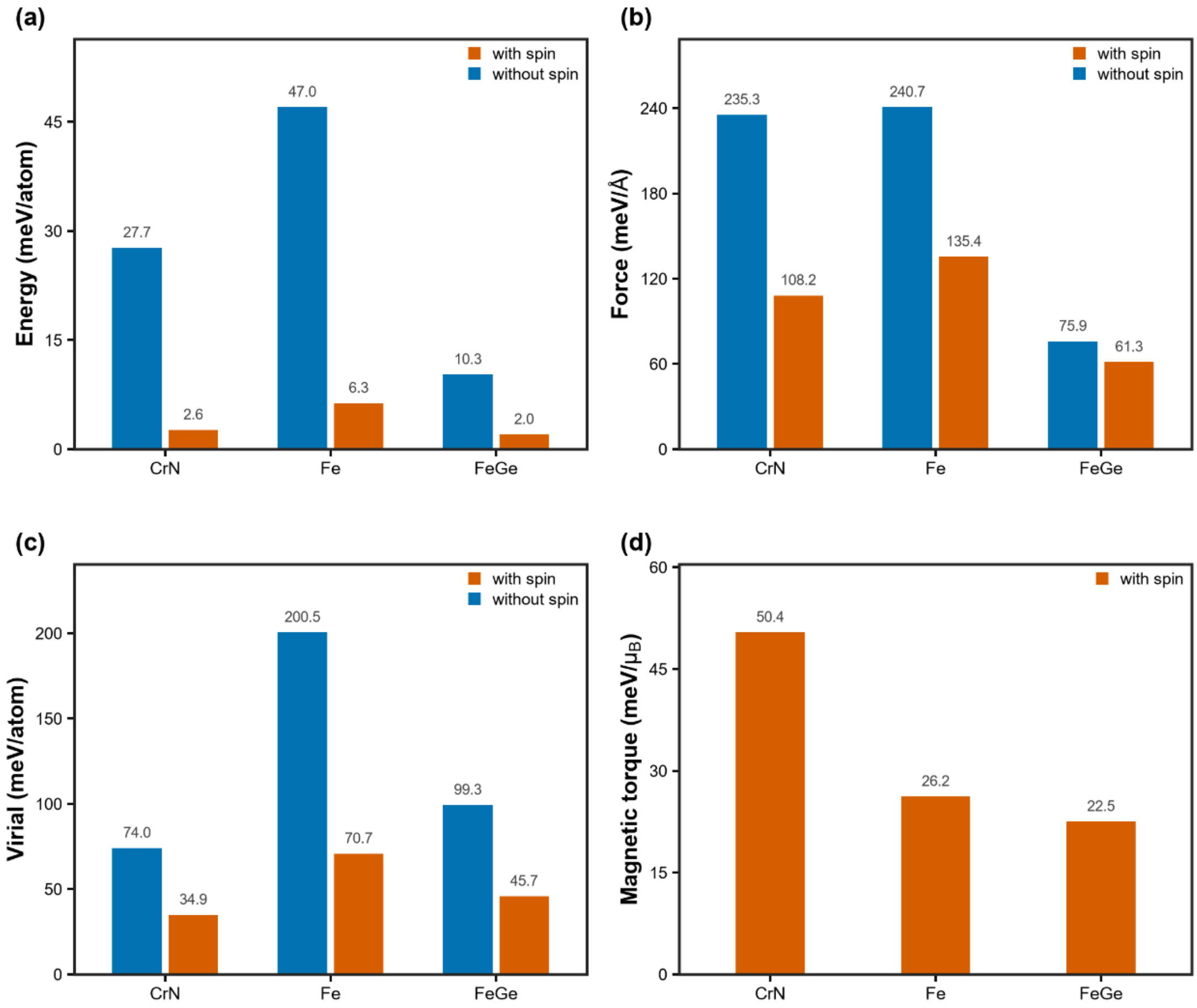

**Fig. 2. Training accuracy of standard NEP and magnetic NEP for CrN with collinear magnetic configurations and Fe and FeGe with noncollinear magnetic configurations.** Panels (a)–(c) compare the root-mean-square errors (RMSEs) in energy, atomic forces, and virials, respectively, for models trained on the same dataset for each system using the same number of optimization iterations. Panel (d) shows the evolution of the magnetic-force RMSE during magnetic NEP training, where the magnetic force is defined as $-\partial U/\partial \mathbf{S}_i$. Energy and virial RMSEs are reported in meV/atom, force RMSEs in meV/Å, and magnetic-force RMSEs in meV/$\mu_{\mathrm{B}}$.

To evaluate the role of the magnetic descriptors, we trained standard NEP models without magnetic descriptors and extended NEP models with magnetic descriptors for CrN with collinear magnetic configurations and Fe and FeGe with noncollinear magnetic configurations. For each system, both models were trained using the same training dataset and the same number of optimization iterations. Figure 2(a)-(c) compares the root-mean-square errors (RMSEs) in energy, atomic forces, and virials. For CrN, Fe, and FeGe, respectively, standard NEP yields energy RMSEs of 27.7, 47.0, and 10.3 meV/atom, force RMSEs of 235.3, 240.7, and 75.9 meV/Å, and virial RMSEs of 74.0, 200.5, and 99.3 meV/atom. Incorporating magnetic descriptors reduces the corresponding energy RMSEs to 2.6, 6.3, and 2.0 meV/atom, force RMSEs to 108.2, 135.4, and 61.3 meV/Å, and virial RMSEs to 34.9, 70.7, and 45.7 meV/atom. The extended NEP consistently outperforms standard NEP across all three systems and all evaluated quantities. The improvement is particularly pronounced for energy predictions, with the energy RMSE reduced by up to approximately one order of magnitude.

These differences highlight the importance of explicitly incorporating magnetic degrees of freedom. Standard NEP uses only atomic positions and chemical species as inputs and therefore cannot distinguish different magnetic configurations for a given atomic structure. Such configurations can nevertheless exhibit different energies, atomic forces, and virials, preventing a model based solely on structural information from establishing a consistent mapping. By introducing magnetic descriptors, the extended model can distinguish these configurations and more accurately capture the joint dependence of the energy and its derivatives on atomic structures and magnetic degrees of freedom. The consistent error reductions across the three systems demonstrate that the proposed descriptors effectively encode magnetic-configuration-dependent information. Figure 2(d) further shows the evolution of the magnetic-force RMSE during training. The final RMSEs for CrN, Fe, and FeGe are 50.4, 26.2, and 22.5 meV/$\mu_B$, respectively, further demonstrating the framework's ability to reproduce magnetic energy gradients.

**Results and Discussion**

Following model training and accuracy assessment, we further evaluate the ability of magnetic NEP to describe CrN, Fe, and FeGe through structural optimization and calculations of physical properties. We first optimize the structures of the three systems using magnetic NEP and compare the equilibrium lattice constants in their respective magnetic states with experimental values, as summarized in Table 1. For antiferromagnetic (AFM) CrN, the optimized lattice constant is 4.1345 Å, in close agreement with the experimental value of 4.1467 Å. For ferromagnetic (FM) Fe and FeGe, the optimized lattice constants are 2.8233 Å and 4.7954 Å, respectively, deviating from the experimental values of 2.8665 Å and 4.7000 Å by approximately 1.51% and 2.03%. In addition, by constructing a commensurate $4 \times 4 \times 146$ B20 supercell that matches the experimentally observed helical pitch of $\sim 70\,\mathrm{nm}$ , structural optimization in the helical magnetic (HM) state yields a folded lattice constant of $4.7080\,\mathrm{Å}$ , in excellent agreement with experiment. Subsequent calculations for FeGe are nonetheless performed within the FM state. Overall, the predicted lattice constants agree well with experimental measurements, indicating that magnetic NEP provides a reasonable description of the equilibrium structures of all three systems in their respective magnetic states. These optimized structures provide a foundation for subsequent calculations and validation of physical properties, including phonon spectra, magnon spectra, and magnetic transition temperatures.

**Table 1.** Comparison of equilibrium lattice constants predicted by magnetic NEP with experimental values for antiferromagnetic (AFM) CrN and ferromagnetic (FM) Fe and FeGe.

| Name | State | Lattice constant (Å) | Experiment (Å) | Reference |
|---|---|---|---|---|
| CrN | AFM | 4.1345 | 4.1467 | [30,31] |
| Fe | FM | 2.8233 | 2.8665 | [32–34] |
| FeGe | FM | 4.7954 | 4.7000 | [35] |

Changes in magnetic order can modify interatomic interactions and the corresponding force constants, thereby altering lattice vibrations. For example, in Fe, the transition from the ferromagnetically ordered state to the high-temperature magnetically disordered state is accompanied by phonon softening, reflecting the close connection between magnetic order and lattice dynamics[7,36]. Studies of $MnPSe_3$, $CrI_3$, CrSBr, and CrOCl have also revealed pronounced effects of magnetic order on lattice vibrations[37–41]. Conventional machine-learning potentials, which take only atomic positions and chemical species as inputs without explicitly incorporating magnetic degrees of freedom, cannot distinguish different magnetic states for the same atomic configuration. Consequently, they cannot capture magnetic-order-dependent lattice dynamics within a unified model. In contrast, magnetic NEP explicitly incorporates the dependence of interatomic interactions on magnetic configurations through magnetic descriptors, providing a means of describing lattice vibrations in different magnetic states.

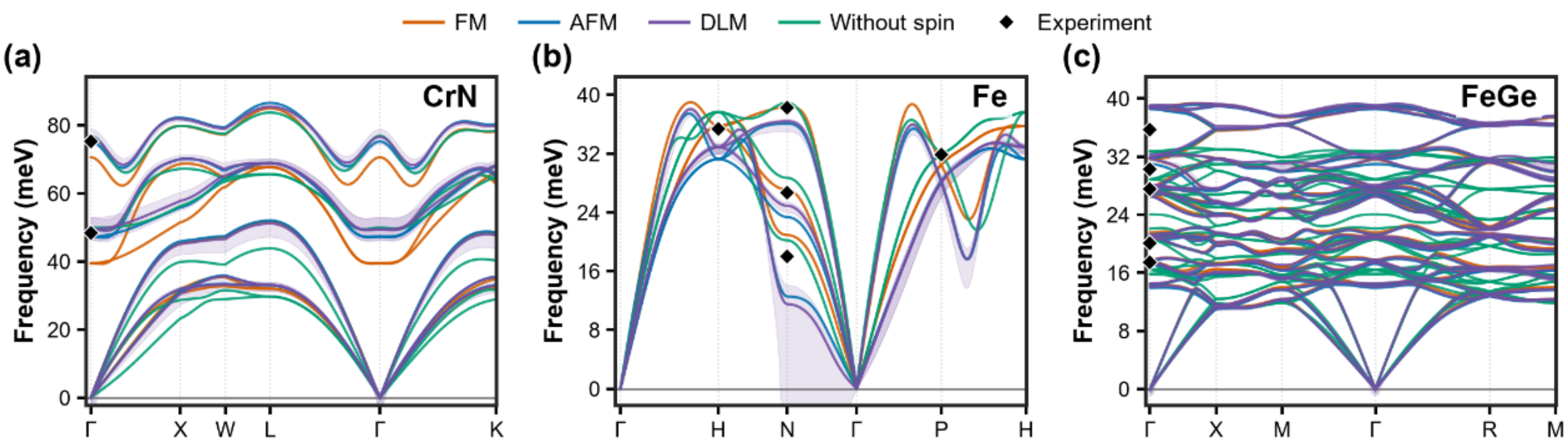

**Fig. 3. Magnetic-configuration-dependent phonon spectra of CrN, Fe, and FeGe calculated using magnetic NEP.** Results for AFM, FM, and magnetically disordered configurations are compared with standard NEP predictions and available experimental data[32,42,43]. Magnetically disordered configurations are constructed using the disordered local moment (DLM) approach. The shaded area denotes the spread across 120 DLM configurations. The experimental reference data for FeGe correspond to characteristic Raman frequencies rather than a complete phonon dispersion.

To assess this capability, we calculate the phonon spectra of CrN, Fe, and FeGe in AFM, FM, and magnetically disordered configurations using magnetic NEP, as shown in Fig. 3. The magnetically disordered states are constructed using the disordered local moment (DLM) approach[44]. All three systems exhibit magnetic-configuration-dependent phonon spectra, whereas standard NEP, which lacks magnetic descriptors, cannot distinguish the phonon responses of different magnetic configurations at the same atomic structure. For CrN, the calculated phonon spectrum in the magnetically disordered state agrees well with experimental measurements for the high-temperature paramagnetic phase. For Fe, the FM phonon spectrum is consistent with experiment, while the transition from the FM-ordered state to the magnetically disordered state produces pronounced phonon softening, particularly along the P-H path. For FeGe, some discrepancies remain between the calculated results and the experimental reference data. Notably, these reference data consist of characteristic Raman peaks used for phase identification rather than a complete phonon dispersion measured along high-symmetry paths. Moreover, the measurement temperature and corresponding magnetic state are not explicitly specified, limiting a direct comparison. Nevertheless, magnetic NEP distinguishes the phonon spectra associated with different magnetic configurations, demonstrating its ability to describe magnetic-order-dependent lattice dynamics.

Beyond the lattice vibrational response to changes in magnetic order, magnons, as collective spin excitations, provide another important test of the model's ability to describe magnetic dynamics. Conventional linear spin-wave theory (LSWT) typically describes magnon dispersion within a linear approximation around a prescribed equilibrium magnetic configuration on a fixed lattice, providing a reference spectrum without interaction-induced broadening. In contrast, neutron-scattering measurements are generally performed at finite temperatures, where magnetic excitation spectra may be affected by thermal spin fluctuations, lattice vibrations, and their coupling. Conventional fixed-lattice LSWT therefore cannot fully capture finite-temperature spectral renormalization and broadening. These limitations motivate the use of spin-lattice dynamics to investigate the temperature dependence of magnetic excitations.

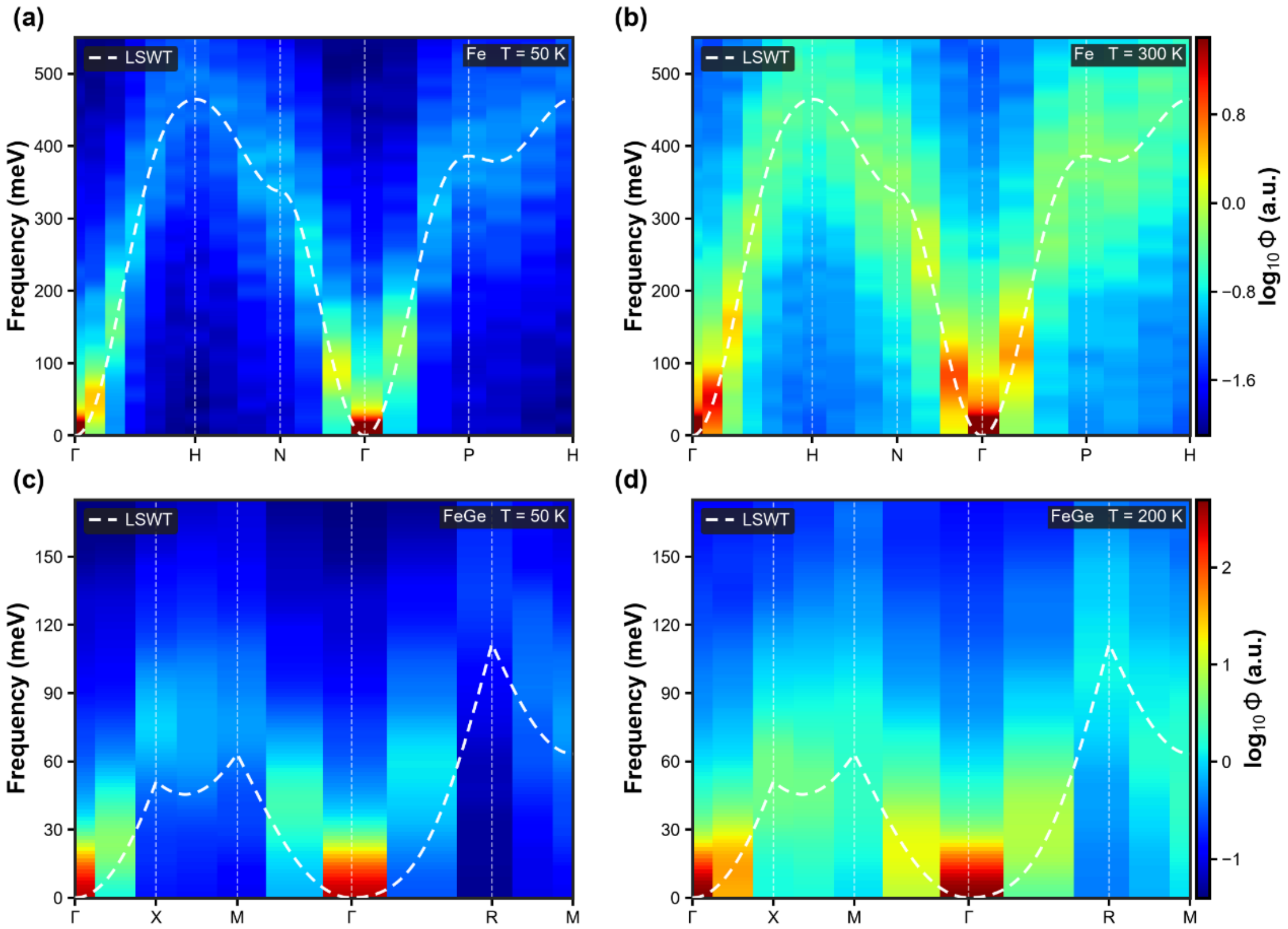

**Fig. 4. Temperature-dependent magnetic spectral energy density (SED) of Fe and FeGe obtained from spin–lattice dynamics simulations using magnetic NEP, compared with linear spin-wave theory (LSWT) predictions.** Panels (a) and (b) show the results for Fe at 50 and 300 K, respectively, while panels (c) and (d) present those for FeGe at 50 and 200 K. The FeGe results include the spectrum at 200 K. The magnetic SED reveals the temperature evolution of the magnetic excitation spectra, while LSWT provides the fixed-lattice reference dispersion.

To investigate finite-temperature magnetic excitations and their coupling to lattice vibrations, we perform spin-lattice dynamics simulations using magnetic NEP, calculate the magnetic spectral energy density (SED), and compare the results with LSWT predictions. As shown in Fig. 4(a) and (b), the magnetic SED peak positions of Fe at 50 K generally follow the LSWT dispersion, indicating that the low-temperature magnetic excitations remain well described by the linear spin-wave picture. Some spectral broadening is nevertheless observed in the low-frequency region. At 300 K, enhanced thermal fluctuations produce pronounced broadening over a wider frequency range, indicating reduced coherence of the magnetic excitations. Once the effects of numerical resolution and other broadening mechanisms are excluded, a larger intrinsic linewidth generally corresponds to a shorter magnon lifetime. FeGe exhibits a similar temperature-dependent trend. Because magnetic order in FeGe is more sensitive to temperature, its magnetic SED dispersion branches are less clearly resolved at 200 K, indicating a substantial influence of thermal fluctuations on magnetic excitations. These results demonstrate that magnetic NEP can describe not only lattice vibrations in different magnetic configurations but also the temperature evolution of magnetic excitation spectra within a unified spin-lattice dynamics framework.

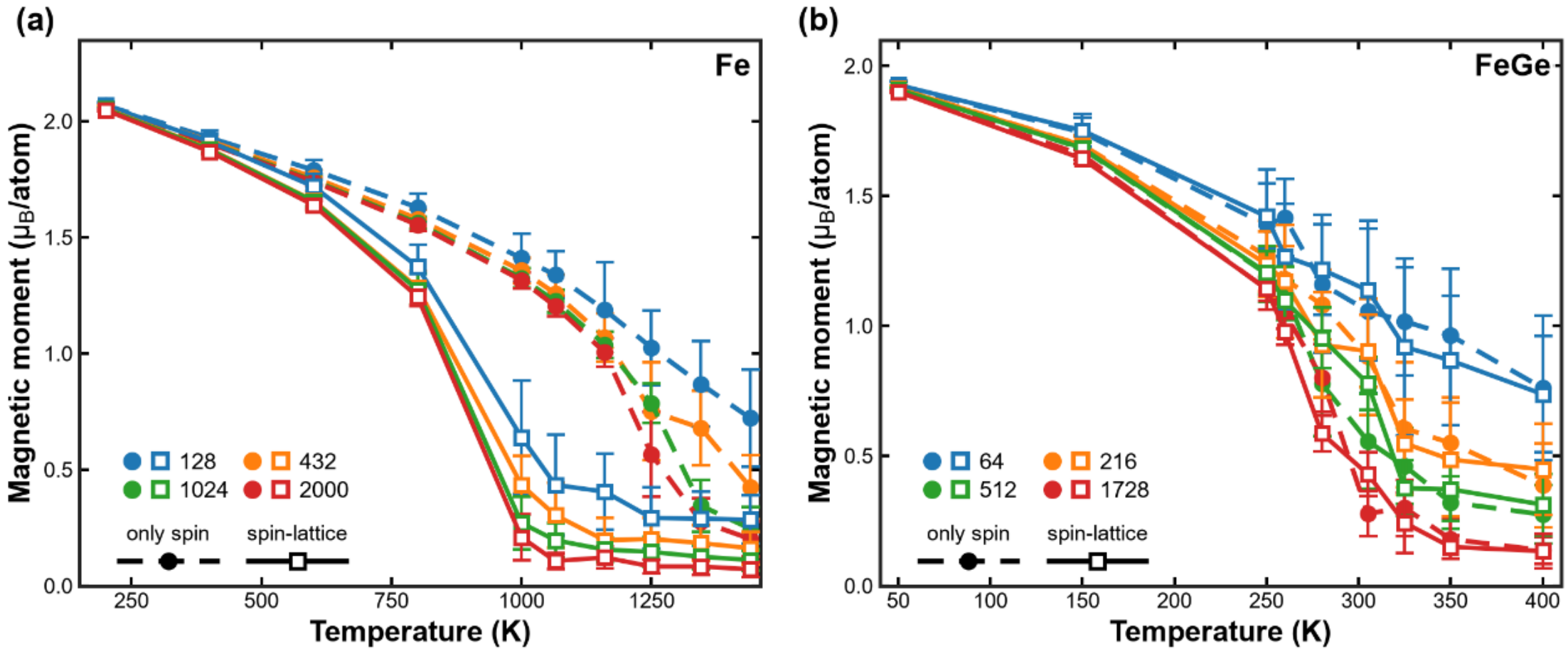

**Fig.5. Effects of spin–lattice coupling and system size on magnetic transitions predicted by magnetic NEP.** (a) Temperature dependence of the average atomic magnetic moment in Fe for different system sizes. (b) Corresponding results for FeGe. Both panels compare fixed-lattice spin dynamics simulations with spin-lattice dynamics simulations that explicitly include spin-lattice coupling.

We further calculate the temperature dependence of the average atomic magnetic moment in Fe and FeGe to assess the ability of magnetic NEP to describe finite-temperature magnetic ordering and magnetic transition temperatures, as shown in Fig. 5. Figures 5(a) and (b) present the results for Fe and FeGe, respectively, at different system sizes. To evaluate the effect of spin-lattice coupling, we employ two simulation schemes: one keeps the atomic positions fixed and evolves only the spin degrees of freedom, while the other evolves both atomic and spin degrees of freedom, explicitly accounting for spin-lattice coupling.

For Fe, both schemes exhibit some system-size dependence, requiring relatively large simulation supercells to obtain well-converged results. At low temperatures, the average atomic magnetic moments obtained using the two schemes differ only slightly; however, the differences become increasingly pronounced as the temperature rises. In the fixed-lattice spin dynamics simulations, the average atomic magnetic moment decreases substantially only at higher temperatures, yielding a higher magnetic transition temperature and a more pronounced system-size dependence. In contrast, including spin-lattice coupling improves the convergence of the temperature-dependent average atomic magnetic moment with respect to system size and yields a magnetic transition temperature of approximately 1000 K, close to the experimental Curie temperature of 1043 K for Fe[33]. These results indicate that spin-lattice coupling has an important influence on finite-temperature magnetic ordering and the magnetic transition temperature of Fe.

For FeGe, the two simulation schemes yield relatively small differences in both the temperature dependence of the average atomic magnetic moment and the magnetic transition temperature. This may be related to the relatively weak lattice thermal vibrations at its lower transition temperature. Furthermore, under the present simulation conditions, including spin-lattice coupling leads to smaller fluctuations in the temperature-dependent average atomic magnetic moment. Overall, these results further demonstrate the ability of magnetic NEP to describe the temperature dependence of the average atomic magnetic moment and magnetic transitions, while highlighting the system-dependent influence of spin-lattice coupling on magnetic transition temperatures.

Finally, we evaluate the computational efficiency of magnetic NEP for large-scale spin-lattice dynamics simulations, with the results summarized in Table 2. The benchmarks are performed on a single consumer-grade GPU using an integration time step of 0.2 fs. The Fe and FeGe systems span a wide range of sizes, from small systems accessible to conventional DFT calculations to large systems containing more than two million atoms. By comparing simulation performance across different system sizes, we assess the computational efficiency of magnetic NEP and its scaling with system size. The results demonstrate that magnetic NEP enables million-atom spin-lattice dynamics simulations on a single consumer-grade GPU, providing a computational tool for investigating magnetic ordering and spin-lattice coupling in large-scale systems.

**Table 2.** Computational performance of magnetic NEP for spin-lattice dynamics simulations of Fe and FeGe at different system sizes on a single consumer-grade GPU. An integration time step of 0.2 fs is used for all benchmarks. The tested systems range from small configurations accessible to conventional density functional theory calculations to large configurations containing more than two million atoms.

| Name | Number of atoms | Speed in ps/day | Time steps (ms) | Memory (MiB) |
|---|---|---|---|---|
| Fe | 16 | 3856.2 | 4.481 | 816 |
| | 128 | 3985.2 | 4.336 | 824 |
| | 432 | 4160.4 | 4.153 | 846 |
| | 1024 | 3562.0 | 4.851 | 822 |
| | 3456 | 3337.0 | 5.178 | 836 |
| | 8192 | 2802.5 | 6.166 | 866 |
| | 27648 | 1243.4 | 13.897 | 966 |
| | 65536 | 584.6 | 29.557 | 1180 |
| | 221184 | 192.4 | 89.802 | 2006 |
| | 524288 | 79.8 | 216.584 | 3574 |
| | 746496 | 56.0 | 308.337 | 4778 |
| | 1362944 | 30.4 | 568.667 | 8018 |
| | 2249728 | 18.3 | 941.732 | 12692 |
| FeGe | 8 | 7048.9 | 2.451 | 816 |
| | 64 | 6144.0 | 2.812 | 820 |
| | 216 | 6231.5 | 2.773 | 830 |
| | 512 | 4929.6 | 3.505 | 818 |
| | 1728 | 4688.7 | 3.685 | 824 |
| | 4096 | 4706.0 | 3.672 | 838 |
| | 13824 | 2941.6 | 5.874 | 892 |
| | 32768 | 1406.2 | 12.289 | 980 |
| | 110592 | 459.5 | 37.602 | 1352 |
| | 262144 | 201.7 | 85.651 | 2044 |
| | 512000 | 104.9 | 164.677 | 3218 |
| | 884736 | 60.9 | 283.589 | 4988 |
| | 1404928 | 38.1 | 453.887 | 7416 |

| | | | |
|---|---|---|---|
| 2097152 | 25.5 | 677.976 | 10592 |
| 2985984 | 17.7 | 976.951 | 14782 |

**Conclusions**

In this work, we incorporate magnetic descriptors into the neuroevolution potential (NEP) to establish a magnetic NEP framework that explicitly accounts for both atomic structures and magnetic degrees of freedom. Using CrN, Fe, and FeGe with collinear or noncollinear magnetic configurations as representative systems, we systematically evaluate its predictive accuracy, ability to describe physical properties, and computational efficiency. Compared with standard NEP, which relies solely on structural information, magnetic NEP distinguishes different magnetic states at the same atomic configuration, substantially reduces prediction errors in energies, atomic forces, and virials, and accurately reproduces magnetic energy gradients. These capabilities provide a foundation for unified dynamical simulations of coupled atomic and spin degrees of freedom.

For physical-property validation, the equilibrium lattice constants predicted by magnetic NEP agree well with experimental values. The model also captures magnetic-configuration-dependent phonon spectra, including phonon softening associated with the transition from ferromagnetic order to magnetic disorder in Fe. Spin-lattice dynamics simulations further reveal temperature-induced spectral broadening and changes in magnon dispersion, extending the description of finite-temperature magnetic dynamics beyond the conventional fixed-lattice linear spin-wave picture. Calculations of the temperature-dependent average atomic magnetic moment show that the influence of spin-lattice coupling on magnetic transitions is system dependent. For Fe, including spin-lattice coupling yields a magnetic transition temperature closer to the experimental value and improves convergence with respect to system size, whereas the two simulation schemes produce relatively small differences in the magnetic transition behavior of FeGe.

While retaining these predictive capabilities, magnetic NEP enables spin-lattice dynamics simulations of systems containing more than two million atoms on a single consumer-grade GPU, demonstrating its capacity for large-scale simulations of magnetic materials. Overall, the framework combines a magnetic-configuration-dependent potential-energy description with efficient dynamical simulations, providing a unified computational tool for investigating finite-temperature magnetic ordering, magnetic excitations, and spin-lattice coupling. It also lays the groundwork for exploring coupled structural and magnetic behavior in more complex magnetic systems.

## Computational Details

### Training Details

The datasets used to train the machine-learning potentials for collinear magnetic CrN[45] and noncollinear magnetic Fe[46] and FeGe[47] were obtained from publicly available datasets released in previous studies. All models were trained exclusively on these datasets, without generating or adding any supplementary training data. Such magnetic datasets are typically constructed using constrained density functional theory (cDFT) calculations to obtain the energies, atomic forces, virials, and magnetic information required for training.

Using these datasets, we trained both standard NEP and magnetic NEP models. For each system, the two models used the same training parameters, differing only in the inclusion of magnetic descriptors and the corresponding magnetic training information. All models employed a feedforward neural network with 30 neurons in the hidden layer and were trained for 100,000 optimization iterations. The radial cutoff radii were set to 5, 7, and 6 Å for CrN, Fe, and FeGe, respectively, while the angular cutoff radii were set to 4, 5, and 5 Å. The angular cutoff radius also served as the cutoff radius for the magnetic descriptors.

**Structural Optimization and Phonon Calculations**

The equilibrium lattice constants were determined by relaxing the unit cells until the atomic forces were below $10^{-3}$ eV/Å. Phonon dispersions were calculated using the finite-displacement method implemented in Phonopy[48], with $4 \times 4 \times 4$, $6 \times 6 \times 6$, and $4 \times 4 \times 4$ supercells for CrN, Fe, and FeGe, respectively. Structural optimizations were initialized from ideal cubic B1-type CrN, body-centered cubic (BCC) Fe, and B20-type FeGe structures. For CrN, a Born-effective-charge-based correction consistent with previous studies was applied, with the diagonal components of the Born effective charge tensors satisfying $Z_{\mathrm{Cr}}^{\alpha\alpha} = -Z_{\mathrm{N}}^{\alpha\alpha} = 4.4$, where $\alpha = x, y, z$[45].

To describe lattice dynamics in the paramagnetic state, we constructed multiple randomly disordered collinear magnetic configurations within the disordered local moment (DLM) approximation and obtained the effective atomic force response by averaging over these configurations[44]. This treatment follows the underlying principle of the spin-space averaging (SSA) approach used in previous calculations of paramagnetic phonons[49]. In SSA, effective atomic forces at finite temperatures are obtained through Boltzmann-weighted averaging of the forces associated with different magnetic configurations, and special quasirandom structures (SQS)[50] can be used to represent magnetic disorder. In this work, we adopted an equal-weight averaging approximation, averaging the atomic forces induced by finite displacements across the randomly disordered magnetic configurations to construct an effective force-constant matrix and calculate the paramagnetic phonon spectrum. This equal-weight treatment corresponds to the approximation that the sampled magnetic configurations occur with equal probability in the infinite-temperature limit.

**Spin–Lattice Dynamics Details**

Spin–lattice dynamics simulations were performed using magnetic NEP as the common potential-energy model for the atomic and spin degrees of freedom. The atomic forces and magnetic forces were evaluated as

$$\mathbf{F}_i = -\nabla_{\mathbf{r}_i} U; \quad \mathbf{w}_i = -\nabla_{\mathbf{S}_i} U, \tag{24}$$

where $U$ is the potential energy, $\mathbf{R}_i$ is the atomic position, and $\mathbf{S}_i$ is the local magnetic moment. Atomic motion was propagated using molecular dynamics, while spin evolution was described by the stochastic Landau-Lifshitz-Gilbert equation[51]. The magnitudes of the local magnetic moments were constrained during spin evolution, allowing their orientations to evolve in response to the instantaneous magnetic forces. Atomic and spin propagation were coupled using a Strang-splitting scheme[52]. A time step of $0.2\ \mathrm{fs}$ was used throughout. The lattice temperature was controlled using a Langevin thermostat with a relaxation time of $100\ \mathrm{fs}$, and the dimensionless spin damping parameter was set to $0.3$. The stochastic magnetic field satisfied the fluctuation-dissipation relation at the target temperature.

To assess the influence of spin-lattice coupling, two simulation schemes were compared. In the fixed-lattice scheme, atomic positions were held fixed and only the spin degrees of freedom evolved. In the coupled scheme, both atomic positions and spins evolved, allowing the atomic forces and magnetic forces to respond to the instantaneous atomic and magnetic configurations. Simulations were performed over a range of temperatures and system sizes to evaluate finite-size effects on magnetic ordering. The average atomic magnetic moment was calculated from the magnitude of the net magnetic moment per atom,

$$\mathbf{m}(T) = < \frac{\sum_i |\mathbf{S}_i(t)|}{N} >_t, \tag{25}$$

where $N$ is the total number of atoms and $<\ldots>_t$ denotes a time average over the sampling interval after equilibration.

**Magnetic Spectral Energy Density Details**

Magnetic excitation spectra were obtained from the spatial and temporal Fourier transforms of transverse spin fluctuations sampled during spin–lattice dynamics simulations, following the general approach used to extract dynamical magnetic spectra from atomistic spin trajectories[7,53]. For Fe, simulations were performed at $50$ and $300\,\mathrm{K}$ using a $6 \times 6 \times 6$ supercell containing $432$ atoms. For FeGe, a $4 \times 4 \times 4$ supercell containing $512$ atoms was used at $50$ and $200\,\mathrm{K}$. Each trajectory was propagated for $20$ ps with a time step of $0.2$ fs, and spin configurations were recorded every $10$ steps, corresponding to a sampling interval of $2$ fs. The initial one-sixth of each trajectory was discarded before spectral analysis.

For each trajectory, two orthogonal unit vectors, $\mathbf{e}_1$ and $\mathbf{e}_2$, perpendicular to the mean magnetization direction were used to define the transverse spin components. Their temporal averages were subtracted to remove the static contribution, and a complex transverse fluctuation was constructed as

$$\psi_j(t) = \delta\mathbf{S}_j(t) \cdot \mathbf{e}_1 + i\delta\mathbf{S}_j(t) \cdot \mathbf{e}_2. \quad (26)$$

The wave-vector-resolved fluctuation amplitude was then evaluated as

$$A(\mathbf{q}, t) = \sum_j \psi_j(t) exp[-i\mathbf{q} \cdot \mathbf{R}_j^0], \quad (27)$$

where $\mathbf{R}_j^0$ denotes the reference position of atom $j$. The magnetic spectral intensity was calculated from

$$I(\mathbf{q}, \omega) \propto < | \int_0^{T_{seg}} w(t) A(\mathbf{q}, t) exp(-i\omega t) dt|^2 >_{seg}, \quad (28)$$

where $w(t)$ is a Hann window and the segment average is taken over trajectory segments. The spectrum was estimated by averaging windowed periodograms of the segments, following the modified-periodogram averaging approach of Welch[54].

The retained trajectory was divided into six segments containing 1251 frames each, corresponding to a nominal segment duration of approximately 2.5 ps and a frequency spacing of approximately 0.40 THz.

**Linear Spin-Wave Theory Details**

Fixed-lattice linear spin-wave theory (LSWT) calculations[55] were performed using the same magnetic NEP models as those employed in the spin-lattice dynamics simulations. Taking a collinear ferromagnetic configuration as the reference state, atomic positions and local magnetic moment magnitudes were held fixed, and an effective Heisenberg model

$$\mathcal{H}_{\text{eff}} = E_0 - \frac{1}{2}\sum_{i \neq j} J_{ij}\, \mathbf{e}_i \cdot \mathbf{e}_j \tag{29}$$

was used to describe exchange interactions in the vicinity of the reference state. Here, $\mathbf{e}_i$ is the unit vector specifying the direction of the local magnetic moment, and $J_{ij}$ is the effective exchange parameter with units of energy. The effective exchange parameters were extracted by applying transverse rotations of $\pm 0.02$ rad to a reference magnetic moment and evaluating the resulting transverse magnetic-force responses at the other sites using a central-difference scheme. For bcc Fe, the average exchange parameters for the first and second nearest-neighbor shells were 32.876 and 14.252 meV, with coordination numbers of 8 and 6, respectively. For B20-type FeGe, the first-nearest-neighbor Fe–Fe exchange parameter had a mean value of 22.090 meV and a within-shell standard deviation of 10.744 meV, with a coordination number of 6; the average second-nearest-neighbor exchange parameter was 1.891 meV. The resulting exchange parameters were then used to calculate fixed-lattice spin-wave dispersions for comparison with the finite-temperature magnetic SED.

**Data availability**

The training datasets used in this work were obtained from publicly available sources. The CrN dataset was obtained from CrN_paramagnetic_AL on GitLab, the Fe dataset from Fe-DeepSpin on AIS Square, and the FeGe dataset from MagPT on Hugging Face.

**Code availability**

The codes used to generate the results presented in this work are available from the corresponding author upon reasonable request. The magnetic NEP implementation and the corresponding molecular dynamics code are publicly available at RongkunChen/GPUMD-spin: Graphics Processing Units Molecular Dynamics for spin methods. The detailed theoretical framework and implementation of spin-lattice dynamics will be reported elsewhere (Li et al., in preparation).

**Acknowledgements**

This research was funded in part by the National Natural Science Foundation of China (Grant No. 12565007), by the Yunnan Fundamental Research Project (Grant Nos. 202601AS070037 and 202401AT070311). Language editing assistance during manuscript preparation was provided by Gemini 3.8 Flash (Google); all contents were critically evaluated and verified by the authors.

**Ethics declarations**

The authors have no conflicts to disclose.